\documentclass[conference]{IEEEtran}
\usepackage{amsmath,amsfonts}
\usepackage{algorithm}
\usepackage{algorithmicx}
\usepackage{array}
\usepackage{textcomp}
\usepackage{stfloats}
\usepackage{float}
\usepackage{url}
\usepackage{verbatim}
\usepackage{graphicx}
\usepackage{svg}
\usepackage{cite}
\usepackage{amsmath}
\usepackage[acronym]{glossaries}
\usepackage{algpseudocode}
\usepackage{cite}
\usepackage{color, soul}
\usepackage[acronym]{glossaries}
\usepackage{lipsum}

\newacronym{IoT}{IoT}{internet of things}
\newacronym{AoI}{AoI}{age of information}
\newacronym{AAoI}{AAoI}{average \gls{AoI}}
\newacronym{AVP}{AVP}{age-violation probability}
\newacronym{DTMC}{DTMC}{discrete time Markov chain}
\newacronym{EH}{EH}{energy harvesting}
\newacronym{ADRA}{ADRA}{age-dependent random access}

\usepackage{color,soul}
\usepackage{graphicx}
\usepackage{amsthm,amssymb}

\begin{document}

\title{Energy and Age-Aware MAC for Low-Power Massive IoT}

 \author{
    \IEEEauthorblockN{Ophelia Giannini\IEEEauthorrefmark{1}, Gabriel Martins de Jesus\IEEEauthorrefmark{2}, Roberto Verdone\IEEEauthorrefmark{1}, Onel Alcaraz L{\'opez}\IEEEauthorrefmark{2}}
    \IEEEauthorblockA{\IEEEauthorrefmark{1}University of Bologna, Bologna, Italy
    \IEEEauthorblockA{\IEEEauthorrefmark{2}Centre for Wireless Communications, University of Oulu, Oulu, Finland 
    \\ophelia.giannini@studio.unibo.it, roberto.verdone@unibo.it}
     \{gabriel.martinsdejesus, onel.alcarazlopez\}@oulu.fi}
}

\maketitle

\begin{abstract}
Efficient multiple access remains a key challenge for emerging Internet of Things (IoT) networks comprising a large set of devices with sporadic activation, thus motivating significant research in the last few years. In this paper, we consider a network wherein IoT sensors capable of energy harvesting (EH) send updates to a central server to monitor the status of the environment or machinery in which they are located. We develop energy-aware ALOHA-like multiple access schemes for such a scenario using the Age of Information (AoI) metric to quantify the freshness of an information packet. The goal is to minimize the average AoI across the entire system while adhering to energy constraints imposed by the EH process. Simulation results show that applying the designed multiple access scheme improves performance from 24$\%$ up to 90$\%$ compared to previously proposed age-dependent protocols by ensuring low average AoI and achieving scalability while simultaneously complying with the energy constraints considered.
\end{abstract}


\section{Introduction}

\IEEEPARstart{T}{he} growing number of devices in \gls{IoT} networks demands efficient deployment and operation. Two critical optimization dimensions are energy usage and timely information updates. Indeed, these devices need a reliable energy supply for sensing the environment, processing the acquired data, executing algorithms, and communicating with other nodes in the network \cite{Lopez:2024}. Meanwhile, the information delivered to the central node or server should be as fresh as possible, which may increase energy consumption, as updates are required more frequently.

In most applications, the devices are not connected to the power grid but rely on batteries. In large networks, replacing depleted batteries is expensive, unsustainable, and time-consuming. A solution to extend the lifetime of batteries is \gls{EH}, a technique for devices to obtain energy from their surroundings, realizing a continuous power supply \cite{Shaikh:2016:RSER}. Adopting \gls{EH} is also a sustainable solution as it reduces the emissions footprint of \gls{IoT} systems \cite{Lopez:2021:IEEE}, increases devices' durability, and reduces the costs and resources needed for the production, deployment, maintenance, and disposal operations of the devices \cite{Lopez:2023:IEEE}.  While \gls{EH} has proved to be a way to extend the lifetime of devices, it requires power management at different levels to be effective \cite{Lopez:2024}, including, for example, energy-aware protocols as energy availability may significantly vary over time.

In this work, we focus on energy-aware MAC protocols for devices with \gls{EH} capabilities. Conventional MAC protocols have been studied from an energy-aware perspective \cite{Famitafreshi:2021:Sensors}, but protocols tailored specifically for \gls{EH}-capable devices are emergent. While these protocols must minimize energy consumption, they should not severely compromise the freshness of the information received at the destination node. Notably, the freshness of information can be quantified by the \gls{AoI} metric by considering the time elapsed since the delivery of the last information from the point of view of a monitoring system \cite{Abbas:2023:CC}. Many time-critical \gls{IoT} applications, such as fire detection in forests or healthcare events predictions require the freshest information available \cite{Kahramam:2024:IOTJ}. In a status update system, a massive number of devices need to access the channel unpredictably. In such a scenario, grant-based access schemes cannot be a solution because their high complexity and the resources wasted for overhead make them highly inefficient and non-applicable to the situation \cite{Munari:2021:BalkanCom}. This is why grant-free ALOHA-like protocols are usually explored.

Recent studies focus on policies aiming to minimize the \gls{AAoI} of \gls{IoT} networks \cite{Chen:2020:INFOCOM, Bacinoglu:2018:ISIT, Arafa:2018:ICC, Arafa:2018:ITA, Ngo:2023:GCC}. In \cite{Chen:2020:INFOCOM}, the authors propose \gls{ADRA}, a protocol to minimize the \gls{AAoI}. At the beginning of the time slot, each device verifies if its age is above a threshold, only then transmitting with a certain probability. The threshold must be high enough to delay the transmission to avoid collisions, but not so high that packets become obsolete. Taming such trade-offs in threshold selection is key to optimizing the system performance. In \cite{Bacinoglu:2018:ISIT}, the authors study the age-energy trade-off with a single \gls{EH} device. In the proposed transmission scheme, transmissions occur if the age of the packet is above a threshold that depends on the device's battery level. The optimal threshold is a monotonically decreasing function of the device's battery level, making it easier to transmit as it increases, avoiding low or empty batteries after transmission. For the case with random \cite{Arafa:2018:ICC} or incremental \cite{Arafa:2018:ITA} battery recharges, \cite{Arafa:2018:ICC} and \cite{Arafa:2018:ITA} indicate that the optimal policy has a \textit{multi-threshold} structure. The authors of \cite{Ngo:2023:GCC} explore the slotted ALOHA protocol on a network with \gls{EH} devices. When a device has an update, it transmits it with a probability that is a function of its battery level. This probability function is optimized for the \gls{AAoI} and the \gls{AVP}, defined as the probability of a packet exceeding the maximum \gls{AoI} allowed by the application's requirement. 


Herein, we propose an energy and age-aware ALOHA-like protocol, linking the issue of having fresh data to the one of the limited energy availability in the battery. Unlike \cite{Chen:2020:INFOCOM}, we consider \gls{EH} capable devices, and the battery levels are taken into account when devices decide to start transmissions. Moreover, we consider a network with multiple devices instead of a single device as in  \cite{Bacinoglu:2018:ISIT, Arafa:2018:ICC, Arafa:2018:ITA}. This study differentiates from \cite{Ngo:2023:GCC} in two aspects: firstly, while they follow a probability-based approach, here we also introduce a threshold; secondly, the system is designed such that devices' batteries are always above a predefined minimum level, meaning they can never deplete completely. This allows devices to keep basic operations alive even in the absence of energy arrivals, increasing the efficiency of the solution and the flexibility in its application in comparison to \cite{Ngo:2023:GCC}. In this paper We consider an \gls{IoT} network composed of several \gls{EH} devices and propose a threshold-based scheme, that takes into account the energy stored in the battery and the current \gls{AoI} of devices to deliver a low \gls{AAoI} performance, significantly reduced compared to \cite{Chen:2020:INFOCOM}. {In our protocol, the current }\gls{AoI} {and battery level of a device influence its decision to access the shared channel as it is used to determine the transmission probability, with weights selected such that the} \gls{AAoI} {of the whole system is minimized. }

The remainder of this paper is organized as follows. In Section \ref{sec:system_model} we introduce our system model and describe the problem formulation. In Section \ref{sec:analytical_formulation} we present our scheme and detail the analytical formulation for the \gls{AoI} and the energy levels of the devices' batteries. In Section \ref{sec:results} we present the simulation results and compare our scheme to previous works. Lastly, in Section \ref{sec:conclusions}, we conclude the paper.

\section{System Model}\label{sec:system_model}
\begin{figure}
    \centering
    \includegraphics[width=\linewidth]{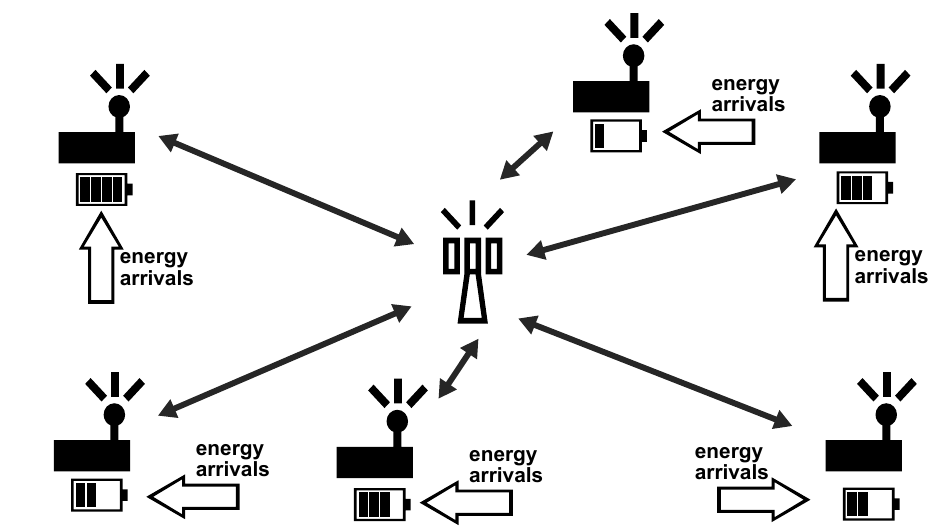}
    \caption{Illustration of the system model considered in this work. Each \gls{IoT} device is equipped with its own battery and is subject to independent energy arrivals. The devices send updates to a central server and wait for acknowledgments.}
    \label{fig:system_model}
\end{figure}

We consider a system composed of $D$ \gls{IoT} devices communicating with a central server through a shared channel. Time is divided into slots and packets have a fixed length and fit inside a single time slot. We assume a collision channel model \cite{Ephremides:1998:TIT} such that a collision occurs when two or more sources transmit in the same time slot, resulting in packet loss. After transmitting, devices receive a feedback message from the receiver if the transmission is successful. Fading channels and noise are neglected, as these are assumed to be resolved at the physical layer. Furthermore, each device keeps track of its own \gls{AoI} {based on the feedback from the receiver and its own internal clock, updating it at the end of each time-slot.} An illustration of the  system model considered is presented in Fig. \ref{fig:system_model}.

\subsection{Energy model}
Each device $i$ has an instantaneous energy level $E_i(t)$. This energy is harvested from the environment and stored in a finite battery of capacity $B$. Energy arrivals $E_i^h(t)$ are independent for each device and follow a Bernoulli distribution adopting values $1$ and $0$ with probability $\eta$ and $1-\eta$, respectively. {If transmitting would cause the battery to be completely depleted, the device must delay the transmission until it has more energy:} at least a quantity of energy $E_\text{min}$ must be stored in the battery after a transmission. This allows the device to carry out basic operations and avoid being shut down. At each transmission, successful or not, a quantity of energy $E$ is used and subtracted from the battery level. Note that $E$ takes into account the energy needed to transmit and to receive an acknowledgment from the central monitor so that the device properly updates the \gls{AoI} status of its data packet. Both $E$ and $E_\text{min}$ depend on the application specifics and are assumed to be the same for each device.
When a transmission occurs the energy level is updated according to
\begin{equation}\label{e_update}
    E_i(t+1) = \min\{E_i(t) + E_i^h(t),B\} - E T_i(t),
\end{equation}
where
\begin{equation}
    T_i(t) = 
    \begin{cases}
    1, & \text{if device $i$ transmits in slot $t$}\\ 
    0, & \text{otherwise}
    \end{cases}.
\end{equation}

\subsection{AoI model}
The \gls{AoI} of each device is an integer value defined as the number of instants passed from the latest packet received correctly on the server side. The time instants of the updates are $T_i^u$, where $u$ refers to the update number. Given the latest update instant, the \gls{AoI} is defined at time $t$ as
\begin{equation}
\Delta_i(t) = t - \max\{T_i^u:T_i^u < t\}.
\end{equation}
The inter-update time is expressed by the integer quantity $X_i^u = T_i^u - T_i^{u-1}$. 
The maximum permissible \gls{AoI}, $\Delta_\text{max}$, depends on the application’s requirements. Herein, we capture application requirements violations via \gls{AVP}, which is the ratio between the number of packets that exceed $\Delta_\text{max}$, and the total number of packets generated $U$, i.e.,
\begin{equation}\label{eq:AVP}
    \text{\gls{AVP}} = \frac{\mathbf{card}(u: \Delta(u) > \Delta_\text{max})}{U},
\end{equation}
where $\Delta(u)$ denotes the \gls{AoI} of packet $u$ immediately upon either being discarded or received.

As devices do not attempt retransmissions of packets, any time they have a packet to transmit, that packet has \gls{AoI} equal to one. When the packet is received, the central monitor updates the \gls{AoI} of the device to the \gls{AoI} of the packet, i.e., to one.  This is why we will refer to the device's \gls{AoI} instead of the packet's. When the transmission is successful, the device's \gls{AoI} is reset, i.e., $\Delta_i(t+1) = 1$. Otherwise, in the time slots the device does not transmit or in case of collision, the device's \gls{AoI} is updated following
\begin{equation}\label{age_update}
    \Delta_i(t+1) =
    \begin{cases}
        \Delta_i(t) +1, & \text{if} \ \Delta_i(t)<\Delta_\text{max} \\
        1, & \text{if} \ \Delta_i(t)=\Delta_\text{max}
    \end{cases}.
\end{equation}
Note that if the \gls{AoI} has already reached its limit $\Delta_\text{max}$, the packet is discarded and, since a new packet is assumed to arrive immediately in the buffer, the device's age is reset. Otherwise, the \gls{AoI} is incremented by 1. The average in time of the \gls{AoI} of a device is computed as
\begin{equation}\label{avg_aoi}
   \bar \Delta_i = \lim_{T \to \infty} \frac{1}{T} \sum_{t=0}^T \Delta_i(t).
\end{equation}
{On the device side, the update of the} \gls{AoI} {is done taking into account the feedback from the receiver. The} \gls{AoI} {is reset when the device receives a positive acknowledgment from the receiver and is increased by one both when it does not receive the acknowledgment and in the time-slots it does not transmit.}

\section{Threshold-based policy}\label{sec:analytical_formulation}
We propose an access policy where devices only start a transmission, with probability $p$, if they have enough energy to stay on after transmitting, and if they jointly reach an age and energy threshold. {The contribution in }\cite{Chen:2020:INFOCOM}{, shown to improve} \gls{AoI}{, inspires this threshold-based approach. Alongside considering the} \gls{AoI} {for transmissions, we use the energy levels of the devices' batteries. This modification improves and extends the previous contribution by never allowing devices to use more energy than they should, and by letting devices with higher battery levels transmit more than those with low energy levels.}

Specifically, we introduce a weight parameter $\alpha$, $ 0\leq \alpha\leq 1$, and compare the weighted sum of functions of the energy level and \gls{AoI} of the device to a threshold $\tau$, $0 \leq \tau \leq 1$. In this policy, devices only know their own update history and the energy arrival distribution. Formally, the device $i$ checks if
\begin{equation}\label{e_cond}
    E_i(t) \geq E+E_\text{min},
\end{equation}
and
\begin{equation}\label{tau_cond}
    (1-\alpha) E^\text{norm}_i(t) + \alpha \Delta^\text{norm}_i(t) \geq \tau,
\end{equation}
where $E^\text{norm}_i(t) = {(E_i(t) - E_\text{min})}/({B-E_\text{min}})$ and $\Delta_i^\text{norm}={\Delta_i(t)}/{\Delta_\text{max}}$
are the normalized energy levels and normalized \gls{AoI} of device $i$, respectively. {In }\eqref{tau_cond}{, the parameters $\alpha$ and $\tau$ must be selected by the operator, and the optimal values vary according to the system parameters, such as the number of devices. In this work, in Section} \ref{sec:results}{, we numerically optimize this value with an exhaustive search, and more formal optimizations are left for future work.} If \eqref{e_cond} and \eqref{tau_cond} are satisfied, the device will start a transmission with probability $p$.
Here, $p$ is introduced to further randomize access transmissions and properly delay transmission instants to minimize collisions. It is modeled as a function of the normalized energy, with a minimum when $E_i(t)=E+E_\text{min}$ and reaching its maximum when the battery is full, i.e. $E_i^\text{norm}(t)=1$.  

\subsection{Transmission probability function}
The probability of transmission must be carefully selected to provide a desirable performance to the network. We consider three possible functions for $p$:
\begin{enumerate}
    \item\label{item:constant} Constant function: $p = K$,
    where $K$ is a constant.
    \item \label{item:linear}Linear function: \begin{equation}
        p = c\left(\frac{(B-E_\text{min})E_i^\text{norm}(t) - E}{B-E_\text{min}-E}\right)
    \end{equation}
    \item \label{item:elliptical}Elliptical function: \begin{equation}
        p = c\left(1-\sqrt{1-\left(\frac{(B-E_\text{min})E_i^\text{norm}(t)}{B-E_\text{min}-E} \right)^2}\right)
    \end{equation}
\end{enumerate}

\begin{figure}[t]
    \centering
    \includegraphics[width=0.45\textwidth]{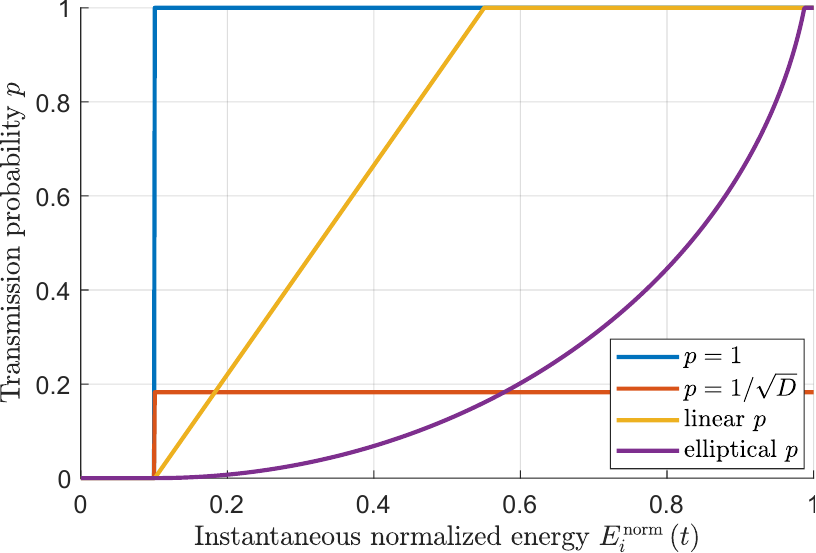}
    \caption{Transmission probability options when $D=30$. The optimal values for $c$ are $c=2.0$ in the linear case and $c=1.2$ in the elliptical case.}
    \label{fig:p_funct}
\end{figure}

These options are illustrated in Fig. \ref{fig:p_funct}. For option \ref{item:constant}), we consider two values, namely, $K=1$ and $K=1/\sqrt{D}$. The value $K=1$ is used to check whether introducing a probability function is beneficial or if it is better to only have the threshold condition. On the other hand,  $K=1/\sqrt{D}$ is chosen considering the ideal case in which there is $1$ transmission out of $p D$, the average number of transmissions in a time slot.

In option \ref{item:linear}), the function is designed to grow linearly with $E_i(t)$ at rate $c$, reaching the maximum value for $p$ at $E_i(t)=B/c + (E_{min}+E)(1-1/c)$. Similarly, option \ref{item:elliptical} is designed to reach maximum when $E_i(t) = \sqrt{1-(1-1/c)^2}(B-E_{min}-E) + E_{min}$. The parameter $c>0$ regulates the maximum $p$, and it is optimized to achieve the minimum \gls{AAoI} depending on the number of devices $D$, {as is the case of parameters $\alpha$ and $\tau$.} This parameter is to be selected such that, when $c<1$, the maximum is $p=c$, while when $c>1$ it $p=1$ is reached before the device has a full battery.  We explore each of these options in Section \ref{sec:results}.

\subsection{DTMC characterization}
\begin{figure}[t]
    \centering
    \includegraphics[width=0.45\textwidth]{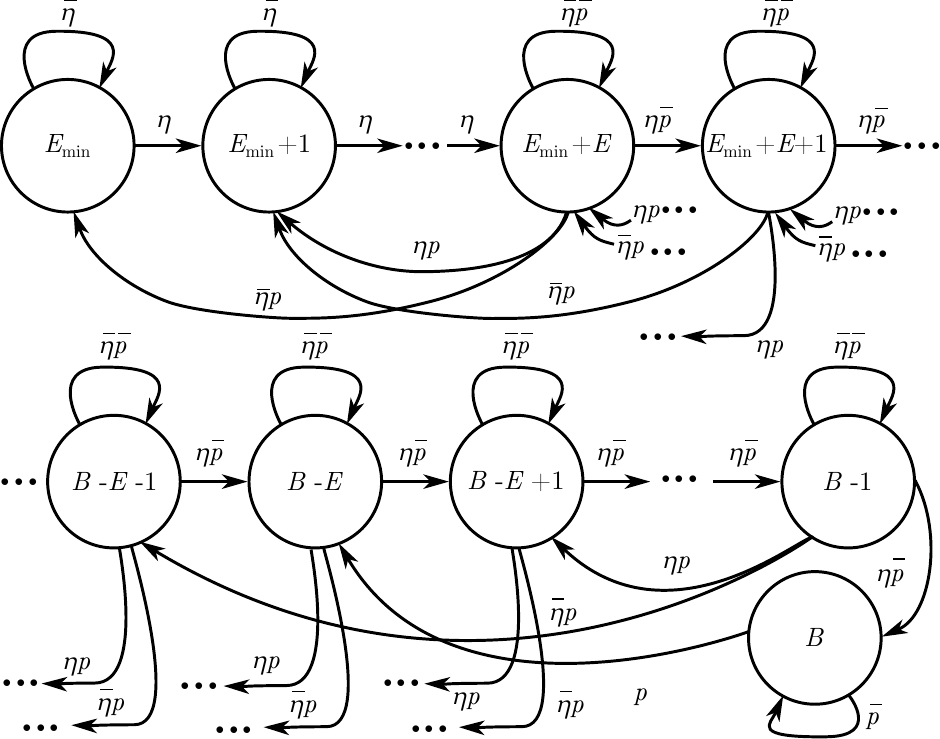}
    \caption{Discrete time Markov chain for energy levels.}
    \label{fig:dtmc_e}
\end{figure}

Herein, we briefly describe the behavior of the system by modeling the \gls{AoI} and energy level of the devices as two linked \gls{DTMC}s, one where the states are the battery levels and one where the states are the \gls{AoI} values.  Note that we do not explicitly specify $p$ to increase tractability.

The \gls{DTMC} for the energy level is illustrated in Fig. \ref{fig:dtmc_e} and is defined by the transition probabilities $P_{m,n}^\text{E}$, which is the probability of the energy's state of the device going from $m$ to $n$. When the \gls{DTMC} of user $i$ is at state $m$ and it does not transmit at time-slot $t$, in the subsequent time-slot, the state can either go to $m+1$ if there is an energy arrival or stay at state $m$ otherwise. Mathematically, 
\begin{align}
    P_{m,m}^\text{E}&=
    \begin{cases}\label{trans_e1}
        1-\eta & E_\text{min} \leq m \leq E_\text{min}+E-1\\
        \bar{\eta}\bar{p} & E_\text{min}+E \leq m \leq B-1\\
        1-p & m=B
    \end{cases},\\
    P_{m,m+1}^\text{E}&=
    \begin{cases}\label{trans_e2}
        \eta & E_\text{min} \leq m \leq E_\text{min}+E-1\\
        \eta\bar{p} & E_\text{min}+E \leq m \leq B-1\\
        0 & m=B
    \end{cases}.
\end{align} 
For both cases, the transition probability depends on $m$. If $m$ is not at least as high as the energy required for transmitting, then the transition probability is readily given by $\eta$, otherwise, the transmission probability $p$ has to be considered. When the battery is full, $\eta$ has no influence, since the device cannot increase its energy level further. Conversely, when the device transmits at time-slot $t$, its next state can be $n = m - E$ or $n=m -E +1$, depending on whether in the same time slot the device also harvests energy. Mathematically,
\begin{align}
    P_{m,m-E}^\text{E}&=
    \begin{cases}\label{trans_e3}
        0 & E_\text{min} \leq i \leq E_\text{min}+E-1\\
        \bar{\eta} p & E_\text{min}+E \leq i \leq B-1\\
        p & i=B
    \end{cases},\\
    P_{m,m-E+1}^\text{E}&=
    \begin{cases}\label{trans_e4}
        0 & E_\text{min} \leq i \leq E_\text{min}+E-1\\
        \eta p & E_\text{min}+E \leq i \leq B-1\\
        0 & i=B
    \end{cases}.
\end{align}
A special case is when $m=B$ so the device cannot capture further energy and can only transmit. From this description of the \gls{DTMC}, one can obtain a stationary distribution $\epsilon_i, i=\{1, 2, \dots, B\}$, of the energy levels. The stationary distribution, however, has a recursive nature and needs to be calculated numerically, and iteratively. For instance, $\epsilon_B = \epsilon_{B-1} {\eta \bar{p}}/{p}$
Meanwhile, 
\begin{equation}
    \epsilon_{i} = \epsilon_{i-1} \frac{\eta \bar{p}}{1-\bar{\eta}\bar{p}},
\end{equation}
for $ i\in[B-E+1, B-1]$. While one can go as far as to obtain
\begin{equation}
    \epsilon_{B} = \epsilon_{B-E} \left(\frac{\eta \bar{p}}{1-\bar{\eta}\bar{p}}\right)^{B-E+1},
\end{equation}
the values of $\epsilon_{B}$ and $\epsilon_{B-1}$, as well as $\epsilon_{B-E-1}$, are required to calculate the value of $\epsilon_{B-E}$, as
\begin{equation}
    \epsilon_{B-E} = \frac{\epsilon_{B-E-1}\eta\bar{p} + \epsilon_{B}p + \epsilon_{B-1}\eta p}{1-\bar{\eta}\bar{p}}.
\end{equation}
This deeply recursive behavior is observed throughout the \gls{DTMC}, thus we limit our analysis to the simulation results in this paper.

\begin{figure}
    \centering
    \includegraphics[width=0.45\textwidth]{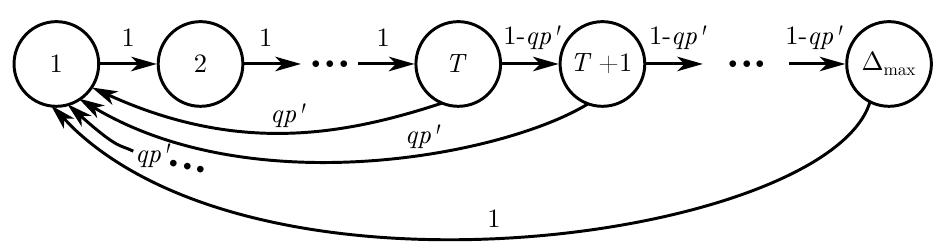}
    \caption{Discrete time Markov chain for \gls{AoI}.}
    \label{fig:dtmc_aoi}
\end{figure}

In the \gls{AoI}'s \gls{DTMC} illustrated in Fig. \ref{fig:dtmc_aoi}, transition probabilities $P_{m,n}^\text{\gls{AoI}}$, which is the probability that the \gls{AoI}'s state of the device goes from $m$ to $n$, are given by 
\begin{align}
    P_{m,m+1}^\text{A}&=
    \begin{cases}\label{aoi_dtmc1}
        1 & 1 \leq i \leq T-1 \\
        1-q p' & T \leq i \leq \Delta_\text{max} \\
        0 & i = \Delta_\text{max}
    \end{cases},\\
    P_{m,1}^\text{A}&=
    \begin{cases}\label{aoi_dtmc2}
        0 & 1 \leq i \leq T-1 \\
        q  p' & T \leq i \leq \Delta_\text{max} \\
        1 & i = \Delta_\text{max}
    \end{cases},
\end{align}
where $p'=p  P_E$ is the overall probability of transmitting, assuming that the \gls{AoI}'s state is higher than $T$, with \begin{equation}
    P_E = \sum_{i=E_\text{min}+E}^B \epsilon_i
\end{equation}being the probability that \eqref{e_cond} is satisfied. Moreover, as other users are symmetric, $q=(1-p')^{D-1}$ is the probability that all the other $D-1$ devices are not transmitting, meaning no collisions, and $T$ represents a ``fictional" threshold on the \gls{AoI}, which is computed as 
\begin{equation}
T=\left\lceil \frac{\tau-(1-\alpha) \bar{E}}{\alpha} \right\rceil,
\end{equation}
and depends on the average energy level $\bar{E}$. This value can be obtained numerically from the energy's \gls{DTMC} as
\begin{equation}
    \bar{E} = \sum_{i=1}^B i \epsilon_i.
\end{equation}

From these definitions, an expression for the stationary probability of the \gls{AoI} can be obtained, which can be used to calculate the network's \gls{AAoI}. Following the derivations from \cite{Chen:2020:INFOCOM}, with the modification that the maximum achievable \gls{AoI} is $\Delta_\text{max}$, we obtain the stationary distribution of the \gls{AoI} as
\begin{equation}\label{stat_aoi}
    \delta_i =
    \begin{cases}
        \delta_{\Delta_\text{max}} + \sum_{k=T}^{\Delta_\text{max}-1}\delta_{k} p'q & i = 1 \leq i \leq T-1\\
        \delta_{i-1} (1-p'q) & T \leq i \leq \Delta_\text{max}
    \end{cases}.
\end{equation}
A closed form of $\delta_i$ can be obtained for the constant $p$ case but becomes unfeasible with other functions for $p$.

\section{Results}\label{sec:results}
We perform Monte-Carlo simulations to compare the performance of different access policies in terms of \gls{AAoI} and \gls{AVP}. We vary the number of devices $D\in[1;500]$, and set $B=100$ energy units, $E=10$ energy units and $E_\text{min}=1$ energy unit. Furthermore, we consider $\eta=0.5$, and the maximum \gls{AoI} to be  $\Delta_\text{max}=200$.
The \gls{AoI} and battery level are initialized with $\Delta_i(0) = 1$ and $E_i(0) = B$, respectively, for each device.

\begin{figure}[t]
    \centering
    \includegraphics[width=0.45\textwidth]{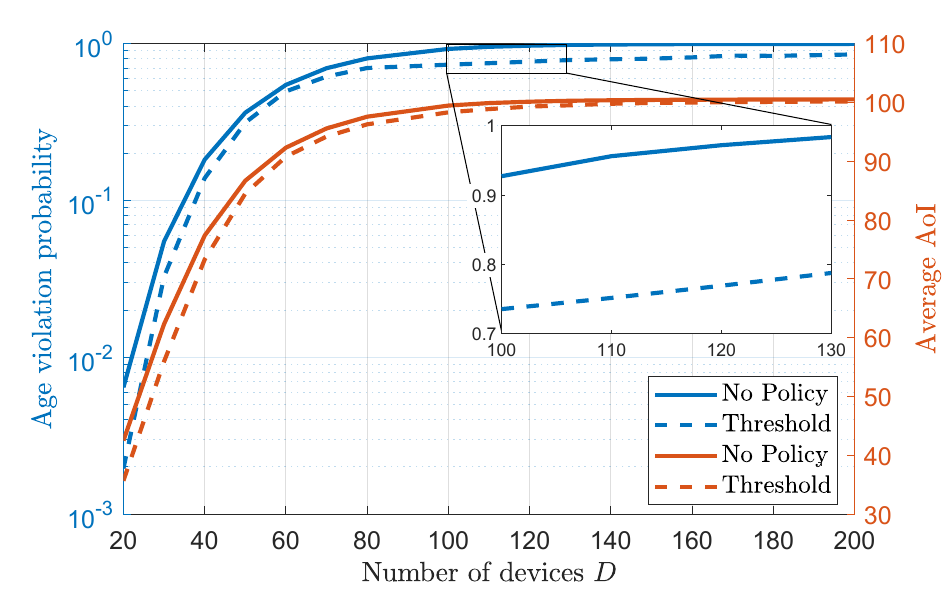}
    \caption{Comparison between applying no policy and applying the threshold-based policy with $p=1$ as a function of $D$.}
    \label{fig:nopolicy}
\end{figure}

\begin{figure}[t]
    \centering
    \includegraphics[width=0.45\textwidth]{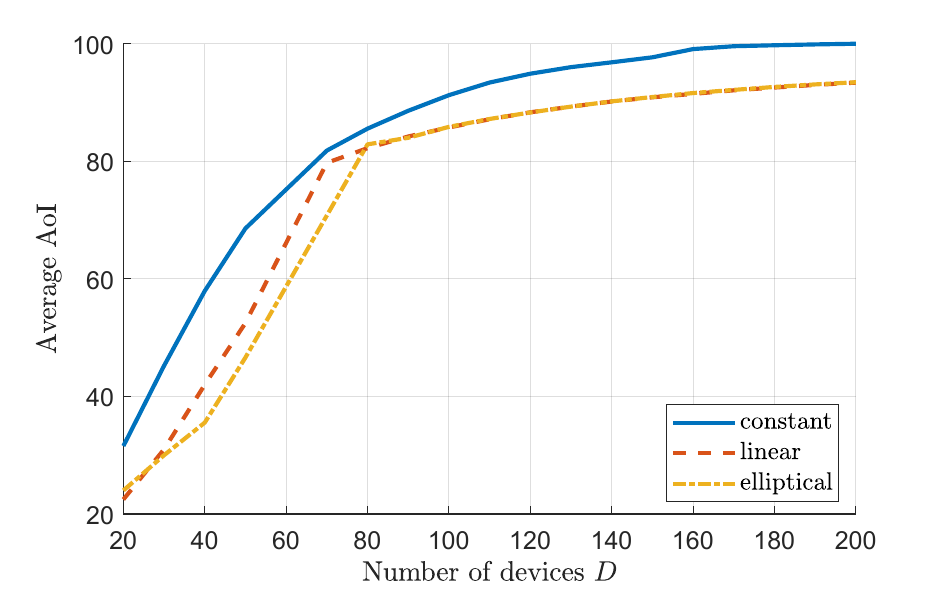}
    \includegraphics[width=0.45\textwidth]{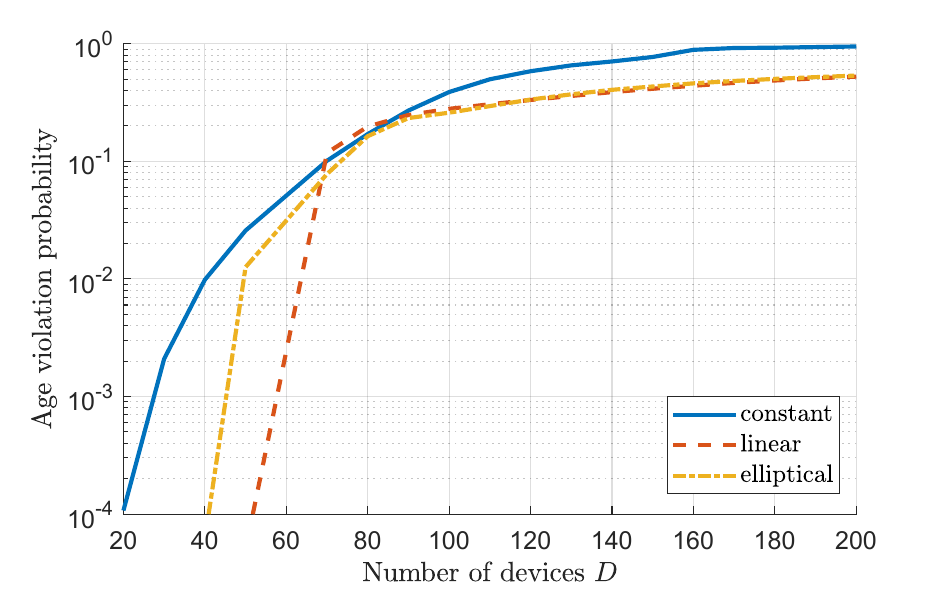}
    \caption{Comparison for  minimum \gls{AAoI} and \gls{AVP} 
    achieved by the different probability functions as a function of $D$, with $p=1/\sqrt{D}$ for the constant case.}
    \label{fig:aoi_avp_th_pfunct}
\end{figure}

We start by comparing the proposed policy's performance with the scenario in which no policy is applied, where devices access the channel as soon as they have enough energy for one transmission, which is equivalent to setting $p=1$ in both cases. We present the results in Fig. \ref{fig:nopolicy}, in which the \textit{threshold only} curve represents the scenario with the transmission probability set to $p=1$. This comparison shows that applying the proposed threshold-based policy leads to a slightly better performance of the systems, particularly when $D<100$.  
As $D$ increases, both strategies lead to similar results in terms of \gls{AAoI}. However, the performance of the threshold-policy, in terms of the \gls{AVP}, is better than when no policy is applied, with a consistent performance gap of around 20\%.

Next, we compare the different probability functions listed in Section \ref{sec:system_model}. We present \gls{AAoI} and \gls{AVP} performance in Fig. \ref{fig:aoi_avp_th_pfunct}, which shows that the minimum \gls{AAoI} is achieved with the elliptical probability function. We optimize the parameter $c$ in the linear and elliptical functions for $p$. For brevity, we only present the optimal values obtained for each value of $D$.  When $D = 50$, the \gls{AAoI} is $68.50$, $52.52$, and $42.19$ for the constant, linear, and elliptical cases, respectively. In this case, compared to constant $p=1/\sqrt{D}$, the elliptical probability function brings an improvement of $38\%$. However, for $D\geq80$, the elliptical and the linear probability functions have similar performances, both outperforming the $p$ constant. The same trends can be observed in the other metrics, with the elliptical function outperforming all other functions for fewer devices, and having results similar to the linear function as $D\geq 80$. 



\begin{figure}
    \centering
    \includegraphics[width=0.45\textwidth]{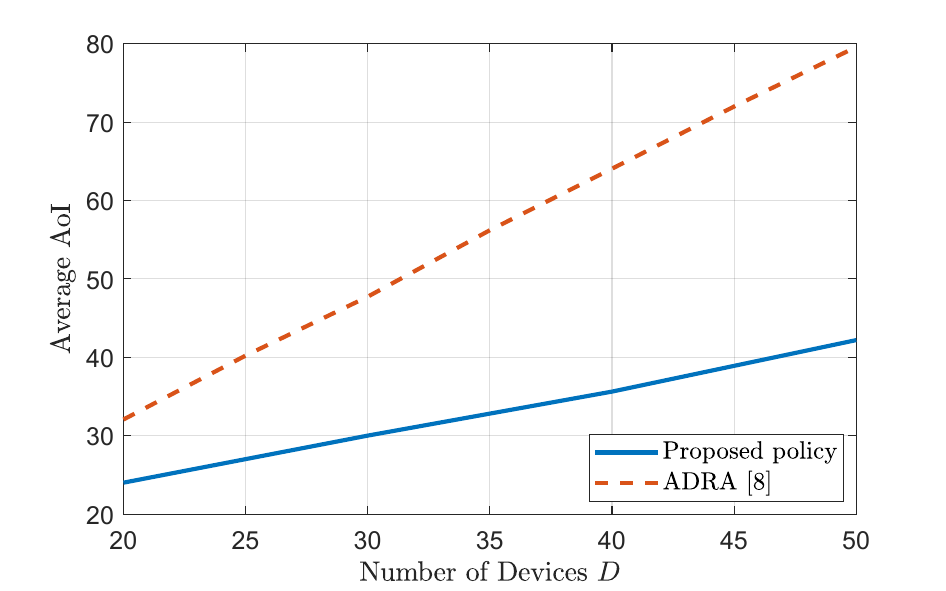}
    \caption{Comparison between policy from \cite{Chen:2020:INFOCOM} and the newly proposed as a function of $D$.}
    \label{fig:chen_comparison}
\end{figure}

Lastly, we compare our proposed policy with the \gls{ADRA} from \cite{Chen:2020:INFOCOM} and present curves of the two policies in Fig. \ref{fig:chen_comparison}. In the \gls{ADRA} policy, the devices transmit with probability $p$ if their \gls{AoI} is above a threshold, without considering the battery level. For a fair comparison, we optimize the age threshold and $p$ in \gls{ADRA} for each point. By using the elliptical function to determine the value of $p$ jointly combining the \gls{AoI} and the battery level, our proposal achieves a better \gls{AoI} performance than that of \cite{Chen:2020:INFOCOM}, while also {complying with the energy usage restrictions}. Moreover, the trends in both curves show that our solution is more scalable, as they grow further apart with $D$.


\section{Conclusions}\label{sec:conclusions}
In this letter, we considered an \gls{IoT} network with \gls{EH} devices and proposed a policy in which they autonomously decide to transmit depending on their \gls{AoI} and battery level. We jointly optimized the protocol's parameters to minimize the \gls{AAoI} and to adapt to the network's dimension. We showed that letting the probability of transmission vary based on the battery level, as well as by considering an age and energy-based threshold for transmissions, can significantly improve the \gls{AoI} performance. The gains were shown to be as high as 60\% when compared to a previously proposed age-dependent scheme.
Future works could focus on extending the system to include physical layer issues, for example, by considering a fading channel and making the energy needed for a transmission $e$ adaptive to channel conditions and to the distance between the monitor and devices.

\section{Acknowledgement}
This work has been funded by the Research Council of Finland (former Academy of Finland) 6G Flagship Programme (Grant Number: 369116), and ECO-LITE (Grant Number: 362782).
\bibliographystyle{IEEEtran}
\bibliography{main}

\end{document}